\documentclass[12pt]{article}
\usepackage{amsmath,amssymb,bm,graphicx}
\usepackage{color} 

\newcommand{\pslash}{\not \! p}

\newcommand{\eslash}{\not \! \epsilon}

\begin{document}

\begin{flushright}
\end{flushright}

\vskip 0.5 truecm

\begin{center}
{\Large{\bf Path integral of neutrino oscillations }}
\end{center}
\vskip .5 truecm
\begin{center}
\bf { Kazuo Fujikawa }
\end{center}

\begin{center}
\vspace*{0.4cm} 
{\it {Interdisciplinary Theoretical and Mathematical Sciences Program,\\
RIKEN, Wako 351-0198, Japan}
}
\end{center}
\makeatletter
\makeatother


\begin{abstract}
We propose an idea of the constrained Feynman amplitude for the scattering of the charged lepton and the virtual W-boson, $l_{\beta} + W_{\rho} \rightarrow l_{\alpha} +  W_{\lambda}$, from which the
conventional Pontecorvo oscillation formula of  relativistic neutrinos is readily obtained using plane waves for all the particles involved.  In a path integral picture, the neutrino propagates forward in time between the production and detection vertices, which are constrained respectively on the 3-dimensional spacelike hypersurfaces separated by a macroscopic positive time $\tau$. The covariant Feynman amplitude is formally recovered if one sums over all possible values of $\tau$ (including negative $\tau$). 
\end{abstract}


\section{Introduction}

It is well-known that the formulation of neutrino oscillations \cite{Pontecorvo, MNS}, if carefully examined,  has some subtleties in the fundamental aspects of quantum mechanics \cite{Beuthe, Akhmedov}. More than one
neutrinos with different masses are produced or detected simultaneously in a quantum mechanical sense and thus the energy-momentum conservation is not obvious. 
There appear two different kinds of neutrino fields; the mass eigenfields $\nu_{k}(x)$, which diagonalize the neutrino mass matrix, and the flavor eigenfields $\nu_{\alpha}(x)$, $(\alpha=e, \mu, \tau)$,  are related to each other by the general mixing formula
\begin{eqnarray}\label{canonical transformation}
\nu_{\alpha}(x) = U^{\alpha k} \nu_{k}(x)
\end{eqnarray}
where $U^{\alpha k}$ stands for the PMNS mixing matrix in a natural extension of the Standard Model. We mainly analyze the Dirac neutrinos in the present paper. This transformation preserves the form of
kinetic terms of neutrinos in the Lagrangian and thus constitutes a
canonical transformation which preserves the canonical anticommutation relations. In the path integral this canonical transformation preserves the
path integral measure invariant. The canonical transformations generally alter the mass terms and interaction terms, and thus they differ from the conventional unitary transformations
of global symmetries and Lorentz transformations in field theory which preserve the form of the Lagrangian invariant. A general class of canonical transformations in connection with neutrinos
are known as the Pauli-Gursey transformation \cite{Pauli} and its generalizations \cite{KF-PG}.

To analyze the basic issues related to the neutrino mixing, we start with a concrete example of the pion decay which provides an initial neutrino in an oscillation experiment
\begin{eqnarray}\label{pion decay}
\pi^{+} \rightarrow \mu^{+} + \nu_{\mu}.
\end{eqnarray}
It is customary to assume that the physical $\nu_{\mu}$ is produced in this decay and then the physical neutrino propagates toward the detector where it is detected by a weak interaction.
 If one of the mass eigenstates in $\nu_{\mu}$ should be identified immediately after the pion decay, for example, such a mass eigenstate due to the reduction of quantum states  would propagate without oscillation, although the charged lepton flavor change $\beta\rightarrow\alpha$ will be induced by the mixing in 
\eqref{canonical transformation}. (The repeated measurements of flavor freedom could also suppress the oscillation.)
 To avoid this difficulty, Kayser  \cite{Kayser} discussed the idea of the wave packets of  particles involved, such as $\mu$ and $\nu_{\mu}$ in \eqref{pion decay}.
This idea of wave packets has become the standard machinery in the analysis of neutrino oscillations and clarified the important aspects of oscillations \cite{Beuthe, Akhmedov, Kayser}. To emphasize the necessity of the wave packets, it is often  stated that the plane waves are unphysical since they are spread in the entire space. 

If one looks at the actual neutrino oscillation experiments, however, the neutrinos are produced by the pion decay and then the neutrinos are detected by a weak interaction inside a huge water Cerenkov detector, for example. These experiments appear to be standard ones common in high energy experiments, and we do not see the particular efforts of experimentalists to generate wave packets in the actual experiments.
In fact, it is very common to use the idea of Feynman diagrams defined by plane waves in almost all the analyses of  the scattering of elementary particles. It is rare to use the wave packets to analyze the scattering of elementary particles.     

One may then wonder if it is possible to describe the neutrino oscillations in terms of the Feynman diagrams using Feynman propagators defined by plane waves for all the particles involved. The main purpose of the present paper is to examine such a possibility. In the explicit analysis of oscillations, we employ the field theoretical formalism, in particular, the path integral.  We do not assume the identification of the physical neutrino immediately after the pion decay  \eqref{pion decay}, and instead the neutrino which appears in the decay is described by the off-shell Feynman propagator which terminates at the weak vertex of the neutrino detector, as in the past field theoretical formulations  \cite{Giunti-Kim1993, Grimus1996, Giunti-Kim1998, Grimus1999} and a related quantum mechanical formulation \cite{Rich}.

We first analyze the Feynman diagram approach to  neutrino oscillations and confirm that the standard covariant Feynman amplitude, as is well known,  does not produce the conventional oscillation formula \cite{Pontecorvo}. We then propose an idea of the constrained Feynman amplitude of neutrino oscillations
using the plane waves for all the particles involved, that readily reproduces the conventional Pontecorvo oscillation formula for relativistic neutrinos. In this scheme, the neutrinos bridge the production and detection vertices located on two 3-dimensional spacelike hypersurfaces, which are defined by two fixed time-slices, fixed $y^{0}$ and fixed $x^{0}=y^{0}+\tau$, separated by a macroscopic time $\tau>0$.  The neutrinos are forced to propagate forward in time and does not propagate  backward in time with negative energy; in this sense, the neutrino propagation is macroscopic and semi-classical in the measurement of oscillations. 
Our proposal is  summarized in \eqref{oscillation formula} and \eqref{formula of neutrino oscillation} below.

\section{The mass and flavor eigenfields of neutrinos in the path integral}

Historically, the quantum mechanical formulation of neutrino oscillations with an emphasis on the Fock space has been discussed by various authors \cite{Blasone, Fujii, Giunti, Tureanu}. This analysis is based on the assumption of the production of the physical neutrino in the pion decay \eqref{pion decay}, for example, and then the physical neutrino thus produced propagates toward the detector in the oscillation experiment. The question is then what kind of vacuum one uses if one assumes the relation \eqref{canonical transformation} in the form 
\begin{eqnarray}
|\nu_{\alpha}\rangle = \sum_{k}U^{\alpha k} |\nu_{k}\rangle
\end{eqnarray}
which, if properly interpreted, is known to lead to the original derivation of Pontecorvo's formula \cite{Pontecorvo}; $\langle\nu_{\beta}(0)|\nu_{\alpha}(t)\rangle \sim \sum_{k}(U^{\beta k})^{\dagger}\exp[-i\sqrt{\vec{p}^{2}+m_{k}^{2}}t]U^{\alpha k}$.

The analyses of the Fock space in neutrino oscillations 
are interesting, and we here comment on 
 the issue of the mass eigenfields and flavor eigenfields in the Fock space formalism from a point of view of the path integral.     
We show that the mass eigenfields and flavor eigenfields are equivalent in defining the  neutrino oscillation amplitudes in the path integral formulation. It is known that only the mass eigenfield is physically relevant in the field theoretical formulation of oscillations \cite{Giunti-Kim1993, Grimus1996}. Nevertheless our simple demonstration of the equivalence of mass eigenfields and flavor eigenfields in the path integral formalism will be interesting.

To analyze the neutrino oscillations, the relevant part of the Lagrangian of a minimal extension of the
Standard Model by adding the right-handed components of neutrinos
and thus assuming the massive Dirac neutrinos is given by
\begin{eqnarray}\label{Dirac Lagrangian1}
{\cal L} &=& \overline{\nu(x)}[ i\gamma^{\mu}\partial_{\mu} - M_{\nu}] \nu(x) + \overline{l(x)}[ i\gamma^{\mu}\partial_{\mu} - M_{l}] l(x)\nonumber\\
&+&  \frac{g}{2\sqrt{2}}\{\overline{l_{\alpha}}\gamma^{\mu}W_{\mu}(x)(1-\gamma_{5})U^{\alpha k}\nu_{k}(x)  + h.c.\}
\end{eqnarray}
where the $M_{\nu}$ and $M_{l}$ stand for the $3\times 3$ diagonalized neutrino and charged lepton mass matrices, respectively, and $U$ stands for the $3\times 3$
PMNS weak mixing matrix. In this representation, the neutrino fields
$\nu_{k}(x)$ correspond to the mass eigenfields. When one integrates over the neutrino variables 
in \eqref{Dirac Lagrangian1} in the path integral, $\int {\cal D}\nu_{k}{\cal D}\overline{\nu_{k}}... \exp\{i\int d^{4}x {\cal L}\}$,  one
obtains
\begin{eqnarray}\label{Dirac Lagrangian2}
\int d^{4}x {\cal L} &=& \int d^{4}x
\overline{l(x)}[i\gamma^{\mu}\partial_{\mu} - M_{l}]l(x)\nonumber\\
&+&\int d^{4}xd^{4}y(\frac{g}{2\sqrt{2}})^{2}
\overline{l^{\alpha}(x)} \gamma^{\lambda}W_{\lambda}(x)(1-\gamma_{5})\nonumber\\
&&\hspace{2cm}\times
U^{\alpha, k}\langle T^{\star}\nu_{k}(x)\overline{\nu_{l}(y)}\rangle(U^{\dagger})^{l,\beta} \gamma^{\rho}W_{\rho}(y)(1-\gamma_{5})l_{\beta}(y)
\end{eqnarray}
where the neutrino propagator $\langle T^{\star}\nu_{k}(x)\overline{\nu_{l}(y)}\rangle$ is defined for the mass eigenfields of Dirac neutrinos
\begin{eqnarray}\label{Dirac propagator}
\langle T^{\star}\nu_{k}(x)\overline{\nu_{l}(y)}\rangle = \int\frac{d^{4}p}{(2\pi)^{4}}\left(
\frac{i}{\pslash -M_{\nu} +i\epsilon}\right)^{kl}
e^{-ip(x-y)}.
\end{eqnarray}
To the accuracy of $O(g^{2})$, the effective vertex
in \eqref{Dirac Lagrangian2}
\begin{eqnarray}\label{scattering amplitude1}
&&(\frac{g}{2\sqrt{2}})^{2}
\overline{l_{\alpha}(x)} \gamma^{\lambda}W_{\lambda}(x)(1-\gamma_{5})\nonumber\\
&&\hspace{1.5cm}\times U^{\alpha, k}\langle T^{\star}\nu_{k}(x)\overline{\nu_{l}(y)}\rangle(U^{\dagger})^{l,\beta} \gamma^{\rho}W_{\rho}(y)(1-\gamma_{5})l_{\beta}(y)
\end{eqnarray}
generates the exact probability amplitude for the  scattering of the charged lepton and the (virtual) W-boson      
\begin{eqnarray}\label{charge lepton scattering} 
l_{\beta} + W_{\rho} \rightarrow l_{\alpha} +  W_{\lambda}
\end{eqnarray}
for the entering charged lepton $l_{\beta}$ and the W-boson at $y^{\mu}$ to the detected outgoing charged lepton $l_{\alpha}$ and the W-boson at $x^{\mu}$ by exchanging the neutrinos. The W-bosons in the above expression are usually
replaced by the hadronic or leptonic charged weak currents, but we use the above amplitude for notational simplicity.  This exact
amplitude describes the charged lepton 
flavor-changing process for specified
$\alpha\neq \beta$ since the basic Lagrangian \eqref{Dirac Lagrangian1} breaks the lepton flavor symmetry, although the fermion number is preserved in the present Dirac neutrinos. The neutrino oscillation is regarded as a very specific charged lepton
flavor-changing process where the conversion rate of the charged leptons oscillates in time or distance between the two vertices $y^{\mu}$ and $x^{\mu}$.

On the other hand, the Lagrangian \eqref{Dirac Lagrangian1} is rewritten in terms of the
flavor eigenfields $\nu_{\alpha}(x)$ defined by $\nu_{\alpha}(x) = U^{\alpha k} \nu_{k}(x)$ in the form
\begin{eqnarray}\label{Dirac Lagrangian3}
{\cal L} &=& \overline{\nu(x)}[ i\gamma^{\mu}\partial_{\mu} - {\cal M}] \nu(x) + \overline{l(x)}[ i\gamma^{\mu}\partial_{\mu} - M_{l}] l(x)\nonumber\\
&+&
\frac{g}{2\sqrt{2}}\{\overline{l_{\alpha}}\gamma^{\mu}W_{\mu}(x)(1-\gamma_{5})\nu_{\alpha}(x)  + h.c.\}
\end{eqnarray}
where the $3\times 3$ mass matrix ${\cal M}$ can be written in the case of the Dirac-type neutrinos as
\begin{eqnarray}
{\cal M} = UM_{\nu}U^{\dagger}.
\end{eqnarray}
We emphasize that the Lagrangian \eqref{Dirac Lagrangian1} is more fundamental than \eqref{Dirac Lagrangian3} in the sense that the derivation of the latter Lagrangian depends on the definition of the mixing matrix in \eqref{Dirac Lagrangian1}, which partly arises from the unitary matrix associated with the mass diagonalization of charged leptons \footnote{The neutrino oscillations can in principle take place even without any mixing among the massive non-degenerate neutrinos in the level of the BEH mechanism. The unitary matrix arising from the diagonalization of the charged lepton mass matrix, which constitutes  the PMNS matrix,  can still cause the  massive neutrino oscillations.}.
One may integrate over the flavor neutrinos in \eqref{Dirac Lagrangian3} in the path integral to obtain
\begin{eqnarray}\label{Dirac Lagrangian4}
\int d^{4}x {\cal L} &=& \int d^{4}x
\overline{l(x)}[i\gamma^{\mu}\partial_{\mu} - M_{l}]l(x)\nonumber\\
&+&\int d^{4}xd^{4}y(\frac{g}{2\sqrt{2}})^{2}
\overline{l_{\alpha}(x)} \gamma^{\lambda}W_{\lambda}(x)(1-\gamma_{5})\nonumber\\
&&\hspace{2cm}\times
\langle T^{\star}\nu_{\alpha}(x)\overline{\nu_{\beta}(y)}\rangle \gamma^{\rho}W_{\rho}(y)(1-\gamma_{5})l_{\beta}(y).
\end{eqnarray}
If one recalls the relation \eqref{canonical transformation}, one has
\begin{eqnarray}\label{propagator conversion}
\langle T^{\star}\nu_{\alpha}(x)\overline{\nu_{\beta}(y)}\rangle= U^{\alpha, k}\langle T^{\star}\nu_{k}(x)\overline{\nu_{l}(y)}\rangle(U^{\dagger})^{l,\beta}
\end{eqnarray}
and thus the Lagrangian \eqref{Dirac Lagrangian4} becomes identical to the Lagrangian \eqref{Dirac Lagrangian2}.

In the framework of the path integral, it is straightforward to derive the relation \eqref{propagator conversion} from the Lagrangian (7) and thus to show the
identical exact scattering amplitudes for two different definitions of neutrino fields. In the
framework of quantum mechanics with an emphasis on the structure
of the Fock space \cite{Blasone, Fujii, Giunti, Tureanu}, however, the direct derivation of the
left-hand side of \eqref{propagator conversion} by defining a suitable vacuum is a major issue. 

We here present a derivation of \eqref{propagator conversion} in the interaction picture
with a more careful definition of the starting propagator and
then using a sum of Feynman diagrams thus defined. We define the
Lagrangian for the neutrino part in \eqref{Dirac Lagrangian3}
\begin{eqnarray}\label{Dirac flavor neutrino}
{\cal L}_{\nu} &=& \overline{\nu(x)}[ i\gamma^{\mu}\partial_{\mu} - {\cal M}] \nu(x)\nonumber\\
&=& {\cal L}_{0} + {\cal L}_{int}
\end{eqnarray}
with $ {\cal L}_{0} = \overline{\nu(x)}[ i\gamma^{\mu}\partial_{\mu}] \nu(x)$ and ${\cal L}_{int} = \overline{\nu(x)}[- {\cal M}] \nu(x)$. We define the
propagator for the flavor field by summing all the Feynman diagrams
defined by the free massless propagator
\begin{eqnarray}\label{massless Dirac}
\langle T^{\star}\nu_{\alpha}(x)\overline{\nu_{\beta}(y)}\rangle_{0} = \int\frac{d^{4}p}{(2\pi)^{4}}\left(
\frac{i}{\pslash +i\epsilon}\right)\delta_{\alpha\beta}
e^{-ip(x-y)}
\end{eqnarray}
which is well-specified by the massless free Lagrangian $ {\cal L}_{0} = \overline{\nu(x)}[ i\gamma^{\mu}\partial_{\mu}] \nu(x)$.

We then obtain the exact propagator defined for the Lagrangian \eqref{Dirac flavor neutrino} by
\begin{eqnarray}\label{Dirac spring}
\langle T^{\star}\nu_{\alpha}(x)\overline{\nu_{\beta}(y)}\rangle &\equiv& 
\langle T^{\star}\nu_{\alpha}(x)\overline{\nu_{\beta}(y)}\rangle_{0}\nonumber\\
&+& \int d^{4}z\langle T^{\star}\nu_{\alpha}(x)\overline{\nu_{\beta^{\prime}}(z)}\rangle_{0}(-i{\cal M})^{\beta^{\prime}\alpha^{\prime}}\langle T^{\star}\nu_{\alpha^{\prime}}(z)\overline{\nu_{\beta}(y)}\rangle_{0} + ..... \nonumber\\
&=&
\int\frac{d^{4}p}{(2\pi)^{4}}\left(
\frac{i}{\pslash -{\cal M} +i\epsilon}\right)^{\alpha\beta}
e^{-ip(x-y)}\nonumber\\
&=&\int\frac{d^{4}p}{(2\pi)^{4}}\left(
\frac{i}{U(\pslash - M_{\nu} +i\epsilon)U^{\dagger}}\right)^{\alpha\beta}
e^{-ip(x-y)}\nonumber\\
&=&\int\frac{d^{4}p}{(2\pi)^{4}}\left(
i(U^{\dagger})^{-1}(\pslash -M_{\nu} +i\epsilon)^{-1}U^{-1}\right)^{\alpha\beta}
e^{-ip(x-y)}\nonumber\\
&=&U^{\alpha k}\int\frac{d^{4}p}{(2\pi)^{4}}\left(
\frac{i}{\pslash - M_{\nu} +i\epsilon}\right)^{k l}
e^{-ip(x-y)}(U^{\dagger})^{l \beta}
\end{eqnarray}
that reproduces \eqref{propagator conversion}.

This conversion of the massless propagator to a massive propagator
by summing a series of \eqref{Dirac spring} is sometimes called  {\em a sum of spring diagrams} since it consists of summing the spring-like Feynman diagrams, and it has been used to formulate a homogeneous renormalization group
equation by Weinberg \cite{Weinberg1}, for example.

The propagator of the mass eigenfields in \eqref{Dirac Lagrangian1}
\begin{eqnarray}
{\cal L}_{\nu} = \overline{\nu(x)}[ i\gamma^{\mu}\partial_{\mu} - M_{\nu}] \nu(x)
\end{eqnarray}
may also be defined by a sum of spring diagrams by defining ${\cal L}_{0} = \overline{\nu(x)}[ i\gamma^{\mu}\partial_{\mu}] \nu(x)$ 
and $ {\cal L}_{int} = \overline{\nu(x)}[- M_{\nu}] \nu(x)$. In this sense, both-hand sides of \eqref{propagator conversion} are on an equal footing. In fact, the canonical transformation of fermion
fields is defined by asking the same form of kinetic terms before and after the transformation and thus characterized by the identical massless
free fermion parts \cite{Pauli, KF-PG}. Thus the above derivation of both hands of
\eqref{propagator conversion} starting with massless fermions is natural in the spirit of canonical
transformations. The equivalence of the mass eigenfields and the flavor eigenfields in the case of Majorana neutrinos shall  be briefly mentioned in Appendix by taking Weinberg's model of Majorana neutrinos \cite{Weinberg2} as an example.

The conversion \eqref{Dirac spring} is analogous to the use of the Bogoliubov transformation in the manner of Nambu-Jona-Lasinio \cite{Nambu}
in the recent paper \cite{Tureanu}, in the sense that the role of  massless fields is emphasized in both cases.

\section{Constrained Feynman amplitude}

 In the field theoretical approach, the “ wave packets” in a broad sense have been used to formulate
the oscillation formula and to clarify
some of the important physical aspects of  neutrino oscillations \cite{Beuthe,  Akhmedov, Giunti-Kim1993, Grimus1996, Giunti-Kim1998, Grimus1999, Rich}.  We here instead propose a simple scheme which works in the description of the neutrino
oscillations using only the plane waves for both internal and external particles.

We re-examine the effective vertex in \eqref{scattering amplitude1}.
We have derived this effective vertex by integrating out the neutrino fields
in \eqref{Dirac Lagrangian1}, but it is more common to encounter this effective vertex 
in the second order perturbation in weak interactions when one analyzes the Feynman
amplitudes. Depending on the final states, the contributions of neutral current couplings leads to a slightly more involved formula \cite{Grimus1996}, but we forgo the analysis of the complications here.

If one integrates over the four-dimensional space-time both at $x^{\mu}$ and $y^{\mu}$, the energy-momentum conservation is imposed at both  $x^{\mu}$ and $y^{\mu}$ and thus one has the {\em off-shell} massive neutrino propagators in \eqref{scattering amplitude1} in general and thus it is not clear if oscillations occur, although the flavor change
of charged leptons such as $\mu \rightarrow e$ will generally take place due to the lepton flavor violation (such as the muon number violation) in \eqref{Dirac Lagrangian1}. This is the well-known fact. 

To be more explicit, we have the  amplitude after the amputation of external legs of Feynman diagrams for the charged lepton flavor changing process $l_{\beta} \rightarrow l_{\alpha}$ as in \eqref{charge lepton scattering}, 
\begin{eqnarray}\label{lepton scattering amplitude}
&&\int d^{4}x d^{4}y e^{iP_{f}x}\overline{u}_{\alpha}(p_{f})\eslash(q_{f})U^{\alpha k}\langle T^{\star}\nu_{L k}(x)\overline{\nu_{L l}}(y)\rangle (U^{\dagger})^{l \beta}\eslash(q_{i})u_{\beta}(p_{i}) e^{-iP_{i}y} \nonumber\\
&& \hspace{2cm} + \left(\eslash(q_{i}) \leftrightarrow \eslash(q_{f})\right) \nonumber\\
&=&\int d^{4}x d^{4}y e^{iP_{f}x}\overline{u}_{\alpha}(p_{f})\eslash(q_{f})U^{\alpha k}(\frac{1-\gamma_{5}}{2})\int\frac{d^{4}p}{(2\pi)^{4}}\left(\frac{i\pslash}{p^{2} -M^{2}_{\nu}+i\epsilon}\right)^{kl}e^{-ip(x-y)}\nonumber\\
&&\hspace{1.5cm}\times (U^{\dagger})^{l \beta} \eslash(q_{i})u_{\beta}(p_{i})e^{-iP_{i}y} \ \ + \left(\eslash(q_{i}) \leftrightarrow \eslash(q_{f})\right) \nonumber\\
&=& (2\pi)^{4}\delta^{4}(P_{f}-P_{i})\{\overline{u}_{\alpha}(p_{f})\eslash(q_{f})(\frac{1-\gamma_{5}}{2})\sum_{k}U^{\alpha k}
\frac{i\pslash}{p^{2}-m^{2}_{k} +i\epsilon}(U^{\dagger})^{k \beta}|_{p^{\mu}=P_{i}^{\mu}}\nonumber\\
&&\hspace{11cm}\times\eslash(q_{i})u_{\beta}(p_{i})
\nonumber\\
&&\hspace{0.5cm}+\overline{u}_{\alpha}(p_{f})\eslash(q_{i})(\frac{1-\gamma_{5}}{2})\sum_{k}U^{\alpha k}
\frac{i\pslash}{p^{2}-m^{2}_{k} +i\epsilon}(U^{\dagger})^{k \beta}|_{p^{\mu}=p_{i}^{\mu}-q_{f}^{\mu}}\eslash(q_{f})u_{\beta}(p_{i})\} \nonumber\\
\end{eqnarray}
where $m_{k}$ is the diagonalized mass of the k-th neutrino, and $P_{i}=p_{i}+q_{i}$ and $P_{f}=p_{f}+q_{f}$ are the entering and the outgoing external total four-momenta, respectively,  which are carried by the charged leptons and (virtual) W-bosons.  Note that we assume the plane waves for all the particles involved and take into account the Bose statistics of two virtual W-bosons.
The neutrinos provide a kind of potential force between the scattering charged leptons.  
 We have the kinematical constraint of the four-momentum conservation 
\begin{eqnarray}\label{pion decay2}
 p_{\nu_{\mu}} = p_{i} + q_{i} =P_{i}, \ \ {\rm or} \ \ 
 p_{\nu_{\mu}} = p_{i} - q_{f}
\end{eqnarray}
with the common four-momentum $p_{\nu_{\mu}}$ for all the neutrino mass eigenstates, which generally imply the {\em off-shell} neutrinos. We see no clear indication of the oscillating behavior in \eqref{lepton scattering amplitude}, although we expect the enhanced behavior near
\begin{eqnarray}\label{onshell condition}
 p_{\nu_{\mu}}^{2}=m_{k}^{2}
 \end{eqnarray}
with k=1, 2, 3. It is important that  we do not have  constraints which would arise  if one should constrain the neutrinos on-shell as in \eqref{onshell condition} \cite{Akhmedov}.
We emphasize that the configurations of the initial and final states consisting of the charged leptons and the (virtual) W-bosons in the present process can be very close to those of the oscillation experiments. But the distance or time scale which characterizes the neutrino oscillation is missing in the formula \eqref{lepton scattering amplitude}, and thus 
the conventional covariant Feynman amplitude does not describe the phenomenon of neutrino oscillations, as is well known.

The neutrino oscillation phenomenon may be regarded as a macroscopic quantum effect, and thus we {\em propose} to generalize the notion of the effective vertex which generates  Feynman amplitudes by
\begin{eqnarray}\label{oscillation formula}
&&\int d^{4}xd^{4}y\delta(x^{0}-y^{0}-\tau)(\frac{g}{\sqrt{2}})^{2}
\overline{l^{\alpha}(x)} \gamma^{\lambda}W_{\lambda}(x)\nonumber\\
&& \hspace{5cm}\times U^{\alpha k}
\langle T^{\star}\nu_{L k}(x)\overline{\nu_{L l}(y)}\rangle(U^{\dagger})^{l\beta} \gamma^{\rho}W_{\rho}(y)l_{\beta}(y)
\end{eqnarray}
with an extra $\delta$-functional constraint $\delta(x^{0}-y^{0}-\tau)$ using a fixed positive macroscopic $\tau$. We call this amplitude with
\begin{eqnarray}
x^{0} - y^{0}  = \tau > 0
\end{eqnarray}
 a {\em constrained Feynman amplitude}. From a point of view of path integral, we sum those “paths” of neutrinos starting on the spacelike hypersurface defined by the fixed $y^{0}$ and ending at the spacelike hypersurface defined by the fixed $x^{0}= y^{0} +\tau$ with a separation by a {\em macroscopic constant} $\tau$, and at the end we sum over $y^{0}$. 
The vertex $x^{\mu}$ of the neutrino propagator is always after the vertex $y^{\mu}$ by a time lapse $\tau >0$. Physics-wise this means that we include the motion of the neutrino propagating forward in time but we
do not include the neutrino propagation backward in time in the path integral. Namely,
we do not include the contributions to the scattering amplitude from
“the neutrino with negative energy” relative to the neutrino propagating
forward in time \footnote{The definition of a particle and an antiparticle is by convention. The particle in our terminology corresponds to the neutrino which we naively identify as propagating  in the oscillation experiments.}; in this sense our prescription  is macroscopic and  semi-classical. 

The path integral with the constraint $x^{0} - y^{0}  = \tau$ does not spoil the symmetry under the simultaneous constant
shifts of all the time variables, namely,
\begin{eqnarray}
x ^{0} \rightarrow x^{0} +\delta t, \ \ \ 
y^{0} \rightarrow y^{0} + \delta t.
\end{eqnarray}
Thus the overall energy-conservation of the observed systems consisting of the entering charged lepton and (virtual) W-boson and the outgoing
charged lepton and (virtual) W-boson in \eqref{oscillation formula} is ensured after integration over $y^{0}$ (or time-averaging over $y^{0}$) but the energy conservation on the  neutrinos described by the Feynman propagator  in the intermediate states is not imposed. 
The momentum
conservation is imposed at all the vertices since we integrate or sum
over the spatial coordinates $x^{k}$ and $y^{k}$ inside the spacelike hypersurfaces. We sum  $x^{k}$ and $y^{k}$ in \eqref{oscillation formula} over all the 3-dimensional spaces but in the physical interpretation we still {\em regard} that the covered spaces in the path integral are ``localized'' seen from a macroscopic scale $\tau$.  For example, a gigantic water Cerenkov counter at Kamioka, for example, which is very large by a microscopic scale, is still very small compared to the oscillation length. Our premise is that we can formulate Feynman amplitudes with idealized plane waves and we can incorporate the possible momentum spread arising from a large but finite detector, for example, by smearing the external states of the charged lepton and the virtual W-boson when we define the final scattering amplitude, if necessary. We assume that these corrections are small and forgo this refinement in the present paper.

To be more explicit, we have the amputated oscillation amplitude generated by \eqref{oscillation formula} when written as a matrix element between the states 
$|W(q_{i})\rangle_{i}\otimes |0\rangle_{f}$ and $\langle l_{\alpha}(p_{f}), W(q_{f})|\otimes \langle \bar{l}_{\beta}(p_{i})|$
~\footnote{We are assuming that the Hilbert spaces at the production vertex and at the detection vertex are effectively factored for large $\tau$ since the weak interactions are short ranged.}, which correspond to the case of the pion decay \eqref{pion decay},
\begin{eqnarray}\label{formula of neutrino oscillation}
&&\int 
d^{4}x d^{4}y\delta(x^{0} - y^{0} -\tau)(\frac{g}{\sqrt{2}})^{2}\nonumber\\
&&\hspace{1cm}\times e^{iP_{f}x}\overline{u}_{\alpha}(p_{f})\eslash(q_{f}) U^{\alpha k}
\langle T^{\star}\nu_{L k}(x)\overline{\nu_{L l}(y)}\rangle(U^{\dagger})^{l\beta}\eslash(q_{i})
v_{\beta}(p_{i})e^{-iP_{i}y}\nonumber\\ 
&=&\int 
d^{4}x d^{4}y\delta(x^{0} - y^{0} -\tau)(\frac{g}{\sqrt{2}})^{2} e^{iP_{f}x}\overline{u}_{\alpha}(p_{f})\eslash(q_{f})(\frac{1-\gamma_{5}}{2}) \nonumber\\
 &&\hspace{1cm}\times U^{\alpha k}
\int \frac{d^{4}p}{(2\pi)^{4}}\left(\frac{i\pslash}{p^{2}-M^{2}_{\nu}+i\epsilon}\right)^{kl}
 e^{-ip(x-y)}(U^{\dagger})^{l\beta}\eslash(q_{i})
v_{\beta}(p_{i})e^{-iP_{i}y}\nonumber\\
&=& (2\pi)^{4}\delta^{4}(P_{f}-P_{i})(\frac{g}{\sqrt{2}})^{2}\overline{u}_{\alpha}(p_{f})\eslash(q_{f})(\frac{1-\gamma_{5}}{2}) \nonumber\\
&&\hspace{1cm}\times U^{\alpha k}\int \frac{dp_{0}}{2\pi}\left(\frac{i\pslash}{p^{2}-M^{2}_{\nu}+i\epsilon}\right)^{kl}e^{-ip_{0}\tau +iP_{i}^{0}\tau}(U^{\dagger})^{l\beta}|_{\vec{p}=\vec{P_{i}}}\eslash(q_{i})
v_{\beta}(p_{i})\nonumber\\
&=&(2\pi)^{4}\delta^{4}(P_{f}-P_{i})(\frac{g}{\sqrt{2}})^{2}\overline{u}_{\alpha}(p_{f})\eslash(q_{f})(\frac{1-\gamma_{5}}{2}) \nonumber\\
&&\hspace{1cm}\times \{\sum_{k}
U^{\alpha k}\frac{\pslash}{2p_{0}}e^{-ip_{0}\tau +iP_{i}^{0}\tau}(U^{\dagger})^{k\beta}|_{ p_{0}=\sqrt{\vec{p}^{2}+m^{2}_{k}}, \ \vec{p}=\vec{P_{i}}}\} \eslash(q_{i})
v_{\beta}(p_{i})
\end{eqnarray}
where $P_{i}=q_{i}-p_{i}$ is the entering four-momentum in the case of the pion decay, for example,  and $P_{f}=q_{f}+p_{f}$ is  the outgoing four-momentum which is carried by the charged lepton and (virtual) W-boson. We are assuming that all the particles are expressed by plane waves. The wave functions $v_{\beta}(p_{i})$ and $\bar{u}_{\alpha}(p_{f})$ stand for the external charged leptons such as $\mu^{+}$ and $e$, respectively, and the wave functions $\epsilon_{\mu}(q_{i})$ and $\epsilon_{\mu}(q_{f})$ stand for the (virtual) W-bosons; the initial $\epsilon_{\mu}(q_{i})$ is actually proportional to the derivative of the pion field in the case of the pion decay. 

Since the phase $e^{iP_{i}^{0} \tau}$
is common  to all the massive neutrinos (for example, $P_{i}=p_{\pi}-p_{\mu}$ in the pion decay \eqref{pion decay}), we have
the essential part of the amplitude from \eqref{formula of neutrino oscillation}
\begin{eqnarray}\label{relativistic oscillation} 
 \sum_{k} U^{\alpha k}\frac{i\pslash}{2p_{0}}e^{-ip_{0}\tau}(U^{\dagger})^{k\beta}|_{p_{0}=\sqrt{\vec{p}^{2}+m^{2}_{k}}, \ \ \vec{p}=\vec{P_{i}}}
\end{eqnarray}
which is a field theoretical version of the oscillation amplitude \cite{Giunti-Kim1993, Grimus1996,Giunti-Kim1998,Grimus1999} if one replaces
\begin{eqnarray}\label{light-cone}
\tau = L
\end{eqnarray}
where $L$ is the spatial distance of two vertices \cite{Akhmedov}. For the relativistic neutrinos, it may be natural to assume that the neutrinos mainly propagate along the light-cone between two hypersurfaces separated by $\tau$, which implies \eqref{light-cone}. For the ultra-relativistic neutrinos,  the factor
\begin{eqnarray}
\frac{\pslash}{2p_{0}}=\frac{1}{2}[\gamma^{0}+\gamma^{l}\frac{p_{l}}{p_{0}}]
\end{eqnarray}
 in \eqref{relativistic oscillation} is regarded to be independent of the neutrino masses since $p_{l}/p_{0}=[p_{l}/|\vec{p}|](1 - (1/2) m^{2}_{k}/|\vec{p}|^{2} + ...)\simeq p_{l}/|\vec{p}|$ and thus not essential for the oscillation (this statement is valid also for $\vec{p}=0$). The amplitude \eqref{relativistic oscillation} then contains the well-known oscillating factor in quantum mechanics \cite{Pontecorvo}
\begin{eqnarray}\label{Pontecorvo}
&&\frac{1}{2}\sin2\theta \left( e^{-i\sqrt{\vec{p}^{2}+m_{1}^{2}}L} - e^{-i\sqrt{\vec{p}^{2}+ m_{2}^{2}}L}\right)\nonumber\\
&=&-i\sin2\theta  e^{[-i(\sqrt{\vec{p}^{2}+m_{1}^{2}}+\sqrt{\vec{p}^{2}+ m_{2}^{2}})L/2]}\sin\{\left(\sqrt{\vec{p}^{2}+m_{1}^{2}}-\sqrt{\vec{p}^{2}+ m_{2}^{2}}\right)L/2\}
\end{eqnarray}
for the specific two-flavor case $\mu\rightarrow e$, for example. 

The minimum length $L$ to measure the oscillation is then specified by 
\begin{eqnarray}
\left|\sqrt{\vec{p}^{2}+m_{1}^{2}}-\sqrt{\vec{p}^{2}+ m_{2}^{2}}\right|L/2 \simeq \left|\frac{m_{1}^{2}- m_{2}^{2}}{4|\vec{p}|}\right|L \sim  1
\end{eqnarray}
depending on the mass differences of neutrinos and the momentum $\vec{p}$ of the virtual neutrinos, which is determined by the measured charged lepton and  W-boson system. From a physics point of view, our prescription \eqref{oscillation formula} probes a tiny energy-splitting contained in the neutrino propagator by varying the macroscopic time $\tau$, which is at least of the order of the energy-time uncertainty limit
\begin{eqnarray}\label{energy-time uncertainty}
\tau \times \left|\sqrt{\vec{p}^{2}+m_{1}^{2}}-\sqrt{\vec{p}^{2}+ m_{2}^{2}}\right|/2 \geq \hbar/2,
\end{eqnarray}
in a notation with explicit $\hbar$. It is notable that, if one sums or integrates over all $\tau$ (including negative $\tau$) in \eqref{oscillation formula}, one would formally recover the original covariant effective vertex \eqref{scattering amplitude1} \footnote{This fact may imply that the breaking of the Lorentz symmetry in  \eqref{oscillation formula} and \eqref{formula of neutrino oscillation} is not fatal. }.  In the present formulation \eqref{oscillation formula}, it might be more appropriate to say that we determine the oscillation time or length by measuring $\tau$ or $L$ rather than predicting the oscillation  time or length.

To clarify the physical picture of our prescription, we repeat the analysis of the energy-momentum balance in \eqref{oscillation formula} and  \eqref{formula of neutrino oscillation}. In the context of the explicit example of the pion decay \eqref{pion decay} at the production vertex, we have the momentum conservation
\begin{eqnarray}
 \vec{p}_{\nu_{\mu}} = \vec{p}_{\pi} - \vec{p}_{\mu} = \vec{P}_{i}
\end{eqnarray}
with the common momentum $\vec{p}_{\nu_{\mu}}$ for all the neutrino {\em mass eigenstates}. We do not impose the energy-conservation at the decay vertex (nor at the detection vertex)
\begin{eqnarray}\label{energy non-conservation}
 p^{0}_{\nu_{\mu}} \neq p^{0}_{\pi} - p^{0}_{\mu} =P^{0}_{i},
\end{eqnarray}
but instead we have the on-shell constraints
\begin{eqnarray}
p^{0}_{\nu_{\mu}}|_{k} = \sqrt{\vec{p}^{2}_{\nu_{\mu}} + m^{2}_{k}}
\end{eqnarray}
arising from the $i\epsilon$-prescription with $\tau>0$ of the Feynman propagator in \eqref{formula of neutrino oscillation} for the k-th mass eigenstate of the neutrino with $k=1, 2, 3$. We ensure the overall conservation of the observed energy-momentum  by the factor $(2\pi)^{4}\delta^{4}(P_{f}-P_{i})$ in \eqref{formula of neutrino oscillation}. 
In our picture, the assembled {\em neutrino mass eigenstates} appearing in the Feynman  propagator are in the virtual states, somewhat analogously to the old fashioned perturbation theory, and consistent with the energy-time uncertainty relation \eqref{energy-time uncertainty}.  

\section{Conclusion}
We have shown that the use of the mass eigenfields or the flavor eigenfields is a matter of canonical transformation of field variables in the
path integral and thus causes no essential differences in the definition
of the exact charged lepton and (virtual) W-boson scattering amplitudes.

We then proposed an idea of the constrained effective vertex which generates Feynman amplitudes  
for the neutrino oscillation process. This scheme is based on the neglect of
the neutrino propagating backward in time relative to the neutrino propagating forward in time, namely, the contribution of the neutrino
with negative energy is neglected, besides assuming the macroscopic time separation by $\tau$. This neglect of the backward propagation generally spoils the Lorentz symmetry. It is then shown that the conventional Pontecorvo oscillation formula of relativistic neutrinos is readily obtained using only
the plane waves for all external and internal particles involved. Our prescription \eqref{oscillation formula} is thus a natural field theoretical counterpart of the conventional quantum mechanical derivation of Pontecorvo's formula \cite{Pontecorvo}. The neutrino oscillation amplitude is valid regardless of the use of neutrino flavor eigenfields or mass eigenfields. Our prescription of the constrained Feynman amplitude gives a novel space-time picture of  neutrino oscillations by measuring the covariant Feynman amplitude for each fixed time-interval parameterized by $\tau$, and the covariant Feynman amplitude is formally recovered if one sums over all possible $\tau$ (including negative $\tau$). It is hoped that a very simple prescription of the present formulation may lead to a new insight into neutrino oscillations.
\\
\\

I thank A. Tureanu for informing me the recent interesting development in the Fock space formalism of neutrino oscillations. The present work is supported in part by JSPS KAKENHI (Grant No.18K03633).

\appendix

\section{Majorana neutrinos}
In this appendix we  briefly comment on the connection of  the mass and flavor eigenfields using Weinberg's model of Majorana neutrinos  \cite{Weinberg2}.
The relevant part of the Lagrangian to analyze the neutrino oscillations is given by
\begin{eqnarray}\label{Majorana Lagrangian1}
{\cal L} &=& {\cal L}_{\nu} + \overline{l(x)}[ i\gamma^{\mu}\partial_{\mu} - M_{l}] l(x)\nonumber\\
&+&
\frac{g}{\sqrt{2}}\{\overline{l_{\alpha}}\gamma^{\mu}W_{\mu}(x)U^{\alpha k}\nu_{L k}(x)  + h.c.\}
\end{eqnarray}
where $M_{l}$ stands for the $3\times 3$ diagonalized charged lepton mass matrix and $U$ stands for the $3\times 3$ PMNS weak mixing matrix. In this representation, the chiral mass eigenfields $\nu_{k}(x)$ are described by the model
Lagrangian $ {\cal L}_{\nu}$ of Majorana neutrinos, for which we adopt Weinberg’s
model that is known to describe the essential aspects of various seesaw models of Majorana neutrinos. The model is defined by an effective
hermitian Lagrangian \cite{Weinberg2}
\begin{eqnarray}\label{Weinberg's model}
 {\cal L}_{\nu} &=& \overline{\nu_{L}}(x)i\gamma^{\mu}\partial_{\mu}\nu_{L}(x)
-(1/2) \{\nu_{L}^{T}(x)CM_{\nu}\nu_{L}(x) + h.c.\}\nonumber\\
&=& (1/2) \{\bar{\psi} i\gamma^{\mu}\partial_{\mu}\psi(x) - \bar{\psi}(x)M_{\nu}\psi(x) \}
\end{eqnarray}
where $M_{\nu}$ stands for the $3\times3$ diagonalized neutrino mass matrix and we defined
\begin{eqnarray}\label{Majorana neutrino}
\psi(x) \equiv \nu_{L}(x) +C\overline{\nu_{L}}^{T}(x).
\end{eqnarray}
The field $\psi(x)$ satisfies the classical Majorana condition identically regardless of the choice of $\nu_{L}$,
\begin{eqnarray}\label{classical Majorana}
\psi(x)= C\overline{\psi(x)}^{T}.
\end{eqnarray}
When the charge conjugation operation defined for a chiral fermion by $\nu_{L}(x) \rightarrow C\overline{\nu_{R}}^{T}(x)$ is not a good symmetry, we define the Majorana fermion by \eqref{classical Majorana} together with  the Dirac equation $[ i\gamma^{\mu}\partial_{\mu} - M_{\nu}]\psi(x)=0$. Following the recent analysis  in \cite{Fujikawa2}, for example,
one can then confirm the equivalence of the oscillation amplitude under a canonical transformation $\nu_{L \alpha}(x) = U^{\alpha k}\nu_{L k}(x)$ in the model \eqref{Majorana Lagrangian1} and \eqref{Weinberg's model}. Our
proposed formula \eqref{oscillation formula} for the process \eqref{charge lepton scattering}  is valid for the Majorana neutrinos also with a due care when neutral current effects are included \cite{Grimus1996}.

\end{document}